# LOW-ENERGY CONDUCTIVITY OF SINGLE- AND DOUBLE-LAYER GRAPHENE FROM THE UNCERTAINTY PRINCIPLE

D. Dragoman – Univ. Bucharest, Physics Dept., P.O. Box MG-11, 077125 Bucharest, Romania,

e-mail: danieladragoman@yahoo.com

ABSTRACT:

The minimum conductivity value as well as the linear dependence of conductivity on the charge density near the Dirac point in single- and double-layer graphene is derived from the energy-time uncertainty principle applied to ballistic charge carriers.

INTRODUCTION

One of the most intriguing properties of both single- and double-layer (bilayer) graphene (for a review see [1]) is the finite value of conductivity in the limit of vanishing density of charge carriers that does not lead to any metal-insulator transition at low temperatures [2]. This minimum conductivity value is universal: it is the same for graphene samples with different mobilities and is practically independent of temperature in a broad temperature range. Experiments confirm that the minimum conductivity of both single- and double-layer graphene is equal to $\sigma_{\min} = 4e^2/h$ (or $e^2/h$ per valley per spin) [2-4]. However, most theoretical works predict that the minimum conductivity of double-layer graphene is different from that of single-layer graphene, which is supposed to attain a value $\pi$ times less than the experiments show, i.e. of only $e^2/\pi h$ per valley per spin. This value has been found using several assumptions and methods: it was obtained in ideal wide graphene strips from the calculation of the mode-dependent transmission probability for different quantization conditions of transversal momenta [5-6] (a similar approach for double-layer graphene leads to $\sigma_{\min} = e^2/2h$ per valley per spin [7]), in planar systems with low carrier densities and linear dispersion relations studied with reduced (3+1)-dimensional gauge theories [8], and in disordered degenerate semiconductors by employing either a mean-field theory [9], a technique that involves disorder-averaged propagators [10] (the same technique applied to double-layer graphene yields $\sigma_{\min} = (3/4)e^2/\pi h$ per valley per spin [11]), or a superfield representation of a weakly disordered system of two-dimensional Dirac fermions with a random mass with zero average [12]. Different calculation methods lead sometimes to different results for the same system. For example, for a system of noninteracting fermions it was predicted that $\sigma_{min} = (e^2/h)(j\pi/2)$ when calculated with the Kubo formula and $\sigma_{min} = (e^2/h)(4j/\pi)$ when an alternative definition of the longitudinal conductivity is used; here $j = 1$ for single-layer graphene, and $j = 2$ for double-layer graphene. Moreover, in the disordered graphene case (the density of states at the Dirac point is finite in disordered graphene), the type of disorder is important: in single-layer graphene away from the Dirac point, a linear dependence of conductivity on charge concentration was obtained for strong scatterers, while a logarithmic dependence characterizes weak scatterers [13], the minimum conductivity being $4e^2/\pi h$ for no disorder and weak disorder that preserves one of the chiral symmetries of the ideal-graphene Hamiltonian. In addition, numerical simulations of the finite-size Kubo formula showed that the linear dependence of the conductivity on the carrier concentration near the Dirac point occurs for the screened Coulomb scatterers but not for short-range scatterers [14]. In double-layer graphene, on the other hand, charge transport calculations in the self-consistent Born approximation revealed that the minimum conductivity is $(e^2/h)(24/\pi)$ in the weak-disorder regime, and $(e^2/h)(8/\pi)$ in the strong-disorder limit [15].

A method to obtain the correct minimum conductivity of single-and double-layer graphene is important for a better understanding of the physics in this unusual material. The aim of this paper is to demonstrate that the experimentally confirmed universal value of the minimum conductivity in single- and double-layer graphene can be retrieved from a Landauer-type calculation of the conductivity if, in addition, an energy-time uncertainty relation is applied to the ballistic charge carriers. Moreover, it is shown that near the Dirac point the conductivity depends linearly on the charge carrier concentration for both single- and double-layer graphene, in agreement with the experiments.

CONDUCTIVITY CALCULATION IN SINGLE-LAYER GRAPHENE

Since the universal nature of the minimum conductivity value per valley per spin is similar to that of the conductance step in a two-dimensional electron gas described by the Schrödinger equation and subject to a transverse constriction, we choose a Landauer-type formalism to find the conductivity of the single-layer graphene. The Dirac-like Hamiltonian for low-energy charge carriers in single-layer graphene has the form

$$H = \hbar v_F \boldsymbol{\sigma} \cdot \boldsymbol{k} = \hbar v_F \begin{pmatrix} 0 & k_x - ik_y \\ k_x + ik_y & 0 \end{pmatrix} \tag{1}$$

where $\boldsymbol{\sigma} = (\sigma_x, \sigma_y)$ consists of Pauli matrices, $\boldsymbol{k} = (k_x, k_y)$ is the momentum/wavevector of the charge carriers and $v_F \cong c/300$ is the Fermi velocity (with $c$ the speed of light), the dispersion relation being given by $E = \pm |\hbar v_F k|$.

Let us assume that charge transport occurs along the $x$ direction when an electric potential is applied between two leads separated by a distance $L$. From (1) it follows that at the Dirac point, i.e. for energy $E = 0$, the components $\psi_1$, $\psi_2$ of the spinor wavefunction satisfy the equations [6]

$$\left(i\frac{\partial}{\partial x} - \frac{\partial}{\partial y}\right)\psi_1(x,y) = 0, \quad \left(i\frac{\partial}{\partial x} + \frac{\partial}{\partial y}\right)\psi_2(x,y) = 0 \tag{2}$$

and have solutions of the form $\psi_1(x,y) \propto \exp(ik_{1x}x + ik_y y)$, $\psi_2(x,y) \propto \exp(ik_{2x}x + ik_y y)$, with $k_{1x} = -ik_y$, $k_{2x} = ik_y$. Unlike in [6], we do not consider imaginary wavevectors, since these are ruled out by the band structure of single-layer graphene, which has no energy gap. The only possibility is then $k_y = k_x = 0$ at the Dirac point in single-layer graphene. If the charge carriers in the electrodes have Fermi energy $E_F$ and wavenumber $k_F$, their x component wavevector in the leads being $\pm\sqrt{k_F^2 - k_y^2}$, the transmission coefficient between the leads is $T = \cos^2\phi/[\cosh^2(k_y L) - \sin^2\phi]$ with $\sin\phi = k_y/k_F$ [6], value which becomes $T = 1$ at the Dirac point.

Near the Dirac point the x and y components of the wavevector are related through $k_x = \pm\sqrt{(E/\hbar v_F)^2 - k_y^2}$ and the components of the spinor wavefunction have the solution

$$\psi_1(x) = \exp(ik_y y) \times \begin{cases} \exp(iqx) + r\exp(-iqx), & x \leq 0 \\ a\exp(ik_x x) + b\exp(-ik_x x), & 0 < x < L \\ t\exp(iqx), & x > L \end{cases}$$

$$\psi_2(x) = \exp(ik_y y) \times \begin{cases} \exp(iqx + i\phi) - r\exp(-iqx - i\phi), & x \leq 0 \\ a\exp(ik_x x + i\varphi) - b\exp(-ik_x x - i\varphi), & 0 < x < L \\ t\exp(iqx + i\phi), & x > L \end{cases} \quad (3)$$

with $q = \sqrt{k_F^2 - k_y^2}$ and $\sin\varphi = k_y/(E/\hbar v_F)$. The transmission coefficient

$$T = \frac{\cos^2\phi \cos^2\varphi}{\cos^2\phi \cos^2\varphi \cos^2(k_x L) + (\sin\phi\sin\varphi - 1)^2 \sin^2(k_x L)} \quad (4)$$

is again unity at normal incidence, i.e. for $k_y = 0$ (see also [16]). Numerical simulations [16] show that the transmission coefficient is higher than 0.9 for $\phi$ angles up to 20°, a fact that properly accounts for charge carriers that contribute to the conductance in real devices and that are not exactly normal to the leads. Therefore $T = 1$ is a suitable approximation of real situations (see also the description of the device used for conductance measurements in [4]).

In order to derive the conductivity, we follow the treatment in [17]. More precisely, in single-layer graphene the conductance of charge carriers with energy $E$ and density of states per valley per spin $D(E) = |E|/[2\pi(\hbar v_F)^2]$ [18] is given by

$$G(E) = \frac{e^2}{\hbar} T(E) D(E) \frac{dE}{dk_x}. \tag{5}$$

Normally, at the Dirac point $G$ should vanish since $D(0) = 0$. However, even at low temperatures the energy of ballistic electrons that contribute to the measured current is not fixed. In fact, the measurement process of electron transmission from one lead to another takes a time interval of the order of the electron traversal time, i.e. of the order of $L/v_F$. Since measurement processes cannot distinguish between the four degenerate charge carriers in single-layer graphene, the average of the time interval between successive detections of charge carriers is $\bar{\tau} = L/4v_F$ and the uncertainty in the energy of ballistic charge carriers can be estimated as

$$\Delta E \cong \hbar/\bar{\tau} = 4\hbar v_F / L. \tag{6}$$

For $L = 200$ nm, $\Delta E = 13$ meV. Due to the uncertainty in energy, the measured conductance is an average of (5) over an energy range equal to $\Delta E$, average which for low temperatures and normal incidence, i.e. for $T(E) = 1$ and $dE/dk_x = \hbar v_F$, is given by

$$\overline{G}(E) = \frac{e^2}{h} \frac{1}{\hbar v_F \Delta E} \int_{E-\Delta E/2}^{E+\Delta E/2} |E| \, dE \tag{7}$$

This conductance can be regarded as the conductance per unit width of the graphene strip. The averaging procedure that must be performed in the case of single-layer graphene is not encountered in quantum wires in which electrons obey the Schrödinger equation because in the latter case the density of states does not depend on energy.

The result of the averaging process depends on the energy value around which it is performed. Around the Dirac point $\overline{G}(0) = (e^2 \Delta E / 4h\hbar v_F) = e^2/(Lh)$, which leads to a conductivity value per valley per spin of

$$\sigma(0) = \sigma_{\min} = e^2/h, \tag{8}$$

in agreement with experimental data. If the average is taken away from the Dirac point, around $E > \Delta E/2$, $\overline{G}(E) = (e^2/h)(E/\hbar v_F)$, while for $0 < E < \Delta E/2$, $\overline{G}(E) = (e^2/h)(E^2 + \Delta E^2/4)/\hbar v_F \Delta E$. Note that away from the Dirac point the averaging procedure does not modify the conductivity value; its effect is important only around the Dirac point.

Experiments indicate that the conductivity away from the Dirac point is proportional to the density of carriers, fact that can be justified by our approach. More precisely, if all electrons with energy $0 < E < E_F$ participate at transport, the total conductance away from the Dirac point per valley per spin and per unit width of the graphene flake, $G_{tot}$, is given at low temperatures by

$$G_{tot} = \frac{e^2}{h} \frac{1}{\hbar v_F} \int_0^{E_F} E dE = \frac{e^2}{h} 2\pi \hbar v_F N \tag{9}$$

where $N = \int_0^{E_F} D(E)dE$ is the low-temperature carrier density in single-layer graphene; the carrier density in single-layer graphene can be changed by (in fact, it is directly proportional to) a gate voltage. At high temperatures the energy dependence of the Fermi-Dirac distribution must be explicitly taken into account, but the result in (9) still holds.

The existence of the minimum conductivity value in single-layer graphene and the linear dependence of the conductivity on the carrier density are unique features of charge transport in graphene and are a direct consequence of both the linear dispersion relation and the chiral behavior of the charge carriers.

CONDUCTIVITY CALCULATION IN DOUBLE-LAYER GRAPHENE

The same method of obtaining the minimum conductivity from the time-energy uncertainty principle can be applied also to double-layer graphene. In this case the Hamiltonian is [2]

$$H = -\frac{\hbar^2}{2m} \begin{pmatrix} 0 & (k_x - ik_y)^2 \\ (k_x + ik_y)^2 & 0 \end{pmatrix}, \tag{10}$$

the chiral character of the charge carriers being preserved, so that the wavefunction is again a spinor. The transmission coefficient of charge carriers is once more $T = 1$ for normal incidence, i.e. for $k_y = 0$, the value of the transmission coefficient being close to unity in an angular range of more than 20° around normal incidence [16]. Therefore, as in the previous section, we consider that the experimental data can be approximated by $T = 1$ and the normal incidence situation. Unlike in single-layer graphene, the energy dispersion relation is parabolic: $E = \pm \hbar^2 k_x^2 / 2m$, the charge carriers having a finite mass $m = 0.05 m_0$, with $m_0$ the free electron mass. Due to this parabolic dispersion relation, in bilayer graphene the average transit time of charge carriers between the leads is energy-dependent (or $k_x$-dependent). More precisely, one can estimate the average time interval between successive detection of charge carriers as $\bar{\tau} = Lm/4\hbar k_x$. The uncertainty in the energy (or wavenumber) of charge carriers can in this case be meaningfully established if we write (6) in the form

$$(\hbar^2 k_x \Delta k_x / m)\bar{\tau} \cong \hbar \tag{11}$$

from which we obtain $\Delta k_x = 4/L$. The low-temperature conductance per valley per spin and per unit width should then be expressed as

$$\overline{G}(k_x) = \frac{e^2}{\hbar} \frac{1}{\Delta k_x} \int_{k_x - \Delta k_x/2}^{k_x + \Delta k_x/2} D(k_x) dk_x = \frac{e^2}{h} \frac{1}{\Delta k_x} \int_{k_x - \Delta k_x/2}^{k_x + \Delta k_x/2} |k_x| dk_x \tag{12}$$

where $D(k_x) = |k_x|/2\pi$ is the density of states in bilayer graphene in the **k**-space for $k_y = 0$. As in the previous paragraph, the averaging procedure is essential near the Dirac point, the conductivity per valley per spin at the Dirac point taking the minimum value

$$\sigma(0) = \sigma_{\min} = e^2/h, \tag{13}$$

which is identical to the value for single-layer graphene. This result is in agreement with experiments. Away from the Dirac point, i.e. for $k_x > \Delta k_x / 2$, the averaging procedure does not modify the conductance $\overline{G}(k_x) = (e^2/\hbar)D(k_x)$, while at intermediate values of the $x$ component of the wavevector a parabolic dependence on $k_x$ is expected, as in the previous section. From the form of the conductance away from the Dirac point it follows that the conductivity is proportional to the carrier density, which can be modified by applying a gate voltage. In deriving this result we have used the chiral behavior of charge carriers, which predicts total transmission at normal incidence, and the dispersion relation in bilayer graphene which, although similar to that in common semiconductors that obey the Schrödinger equation, differs from the latter through the absence of an energy gap. The absence of the energy gap in the density of states leads to the manifestation of the energy uncertainty in the immediate vicinity of the Dirac point, the uncertainty principle having no direct influence on conductance away from the Dirac point.

CONCLUSIONS

We have shown that the minimum conductivity in both single- and double-layer graphene takes the experimentally confirmed value if the time-energy uncertainty relation is taken into account. This uncertainty relation does not influence the conductivity away from the Dirac point, and therefore is manifest only for mesoscopic structures with no energy gap. We have also recovered the linear dependence of conductivity on charge density away from the Dirac point. These results are a consequence of the peculiar energy dispersion relation and the chiral behavior of charge carriers in both single- and double-layer graphene.